\documentclass[12pt,english]{article}
\usepackage[T1]{fontenc}
\usepackage[latin1]{inputenc}
\usepackage{geometry}
\makeatletter

\makeatletter

\makeatletter

\usepackage{geometry}

\usepackage{paralist}



\makeatother

\makeatother

\usepackage{babel}
\makeatother
\begin{document}

\title{Response to a critique of the Borexino result in \char`\"{}A new
experimental limit for the stability of the electron\char`\"{} by
H.V. Klapdor-Kleingrothaus, I.V. Krivosheina and I.V. Titkova}

\maketitle
\textbf{H.O.Back$^{a}$, M.Balata$^{b}$, A. de Bari$^{c}$, T. Beau$^{d}$,
A. de Bellefon$^{d}$, G.Bellini$^{e}$, J.Benziger$^{f}$, S.Bonetti$^{e}$,
C.Buck$^{g}$, B.Caccianiga$^{e}$, L.Cadonati$^{f}$, F.Calaprice$^{f}$,
G.Cecchet$^{c}$, M.Chen$^{h}$, A. Di Credico$^{b}$, O.Dadoun$^{d}$,
D.D'Angelo$^{i}$, A.Derbin$^{j,t}$, M.Deutsch$^{k}$, F.Elisei$^{l}$,
A.Etenko$^{m}$, F. von Feilitzsch$^{n}$, R.Fernholz$^{f}$, R.Ford$^{f}$,
D.Franco$^{e}$, B.Freudiger$^{g}$, C.Galbiati$^{f}$, F.Gatti$^{i}$,
S.Gazzana$^{b}$, M.G.Giammarchi$^{e}$, M.Goeger-Neff$^{n}$, A.Goretti$^{b}$,
C.Grieb$^{n}$, T.Hagner$^{n}$, B.Harding$^{f}$, F.X.Hartmann$^{g}$,
G.Heusser$^{g}$, A.Ianni$^{f,u}$, A.M.Ianni$^{f}$, H. de Kerret$^{d}$,
J.Kiko$^{g}$, T.Kirsten$^{g}$, G.Korga$^{e,s}$, G.Korschinek$^{n}$,
Y.Kozlov$^{m}$, D.Kryn$^{d}$, M.Laubenstein$^{b}$, C.Lendvai$^{n}$,
P.Lombardi$^{e}$, I.Machulin$^{m}$, S.Malvezzi$^{e}$, J.Maneira$^{e}$,
I.Manno$^{o}$, G.Manuzio$^{i}$, F.Masetti$^{l}$, A.Martemianov$^{m,r}$,
K.McCarty$^{f}$, E.Meroni$^{e}$, L.Miramonti$^{e}$, M.E.Monzani$^{b,e}$,
P.Musico$^{i}$, H.Neder$^{g}$, L.Niedermeier$^{n}$, L.Oberauer$^{*}$,
M. Obolensky$^{d}$, F.Ortica$^{l}$, M.Pallavicini$^{i}$, L.Papp$^{e,s}$,
L.Perasso$^{e}$, A.Pocar$^{f}$, R.S.Raghavan$^{p}$, G.Ranucci$^{e}$,
W.Rau$^{b,g}$, A.Razeto$^{i}$, E.Resconi$^{i}$, A.Sabelnikov$^{e,r}$,
C.Salvo$^{i}$, R.Scardaoni$^{e}$, S.Schoenert$^{g}$, H.Seidel$^{n}$,
H.Simgen$^{g}$, M.Skorokhvatov$^{m}$, O.Smirnov$^{j}$, A.Sotnikov$^{j}$,
S.Sukhotin$^{m}$, V.Tarasenkov$^{m}$, R.Tartaglia$^{b}$, G.Testera$^{i}$,
D.Vignaud$^{d}$, S.Vitale$^{i}$, R.B.Vogelaar$^{a}$, V.Vyrodov$^{m}$,
M.Wojcik$^{q}$, O.Zaimidoroga$^{j}$, G.Zuzel$^{q}$. }

$\:$

$\:$

$\:$

$a$)Virginia Polytechnique Institute and State Univercity Blacksburg,
VA 24061-0435, Virginia, U.S.A.;

$b$)I.N.F.N Laboratori Nazionali del Gran Sasso, SS 17 bis Km 18+910,
I-67010 Assergi(AQ), Italy;

$c$)Dipartimento di Fisica Nucleare e Teorica Università di Pavia,
Via A. Bassi, 6 I-27100, Pavia, Italy;

$d$)Astroparticule et Cosmologie (APC) (Université Paris 7, CNRS,
CEA, Observatoire de Paris), 10 rue Alice Domon et Léonie Duquet,
F-75205 Paris Cedex 13, France 

$e$)Dipartimento di Fisica Università and I.N.F.N., Milano, Via Celoria,
16 I-20133 Milano, Italy

$f$)Department of Physics, Princeton University, Jadwin Hall, Washington
Rd, Princeton NJ 08544-0708, USA;

$g$)Max-Planck-Institut fur Kernphysik, Postfach 103 980 D-69029
Heidelberg, Germany;

$h$)Dept. of Physics, Queen's University Stirling Hall, Kingston,
Ontario K7L 3N6, Canada;

$i$)Dipartimento di Fisica Università and I.N.F.N. Genova, Via Dodecaneso,
33 I-16146 Genova, Italy;

$j$)Joint Institute for Nuclear Research, Joliot Curie 6, 141980,
Dubna Moscow region, Russia;

$k$)Department of Physics Massachusetts Institute of Technology,
Cambridge, MA 02139, U.S.A.;

$l$)Dipartimento di Chimica Università di Perugia, Via Elce di Sotto,
8 I-06123, Perugia, Italy;

$m$)RRC Kurchatov Institute, Kurchatov Sq.1, 123182 Moscow, Russia;

$n$)Technische Universitat Munchen, James Franck Strasse, E15 D-85747,
Garching, Germany

$o$)KFKI-RMKI, Konkoly Thege ut 29-33 H-1121 Budapest, Hungary;

$p$)Bell Laboratories, Lucent Technologies, Murray Hill, NJ 07974-2070,
U.S.A;

$q$)Institute of Physics, Jagellonian University, Cracow. ul. Reymonta
4. PL-30059 Krakow, Poland;

$r$)on leave of absence from m;

$s$)on leave of absence from $o$;

$t)$on leave of absence from St. Petersburg Nuclear Physics Inst.
- Gatchina, Russia;

$u)$presently at $b$;

$*$) Project manager.

\begin{abstract}
A recently published article by Klapdor-Kleingrothaus \textit{et al.}
\cite{KKKT} critiques the limit on the stability of the electron
obtained by the Borexino collaboration. We respond here to the criticisms
raised by Klapdor-Kleingrothaus and his colleagues, and re-establish
that our result is based on very conservative premises and that the
{}``indication of a signal of 1.4 $\sigma$'' for the decay of the
electron in the $\gamma+\nu$ channel, reported by Klapdor-Kleingrothaus
and colleagues, is excluded by the Borexino result.


\end{abstract}

\section*{Response to criticism of the Borexino result}

The limit on the stability of the electron against $e\rightarrow\gamma+\nu$
decay reported by Klapdor-Kleingrothaus and colleagues~\cite{KKKT},
$\tau>1.22\times10^{26}$\,yr (68\% c.l.), is weaker than the best
current limit of $4.6\times10^{26}$\,yr (90\% c.l.) established
by the Borexino collaboration with the Counting Test Facility detector
(its second version, CTF-II) operated at LNGS~\cite{Borexino}. The
result from Klapdor-Kleingrothaus et al (Ref. \cite{KKKT}, referred
to as the KKKT article hereafter) is also weaker than an earlier limit
set by the DAMA collaboration \cite{DAMA} of $2.0\times10^{26}$
yr (90\% c.l.). While the weaker KKKT limit is still of interest because
of the different technique used, its authors also directly challenge
the Borexino procedure used to obtain its stronger limit (the DAMA
result is not mentioned in this context).

The criticisms raised in KKKT are not supported by a quantitative
analysis. Some points raised by the authors are irrelevant to the
result of interest, while others raise issues which were discussed
in detail in the Borexino article~\cite{Borexino} and missed by
the authors of KKKT. In the following paragraphs we offer a detailed
examination of all the arguments raised in KKKT to criticize the Borexino
result.

\begin{enumerate}
\item \textit{The background whose origin seems not to be fully known has
been parameterized by six parameters - and it has to be assumed to
behave linearly down to low energies}''.\\
 The CTF-II background at low energies is well understood because
it is dominated by the well-known shape of $^{14}$C beta decay. The
assumption of linear behaviour of the underlying background is only
relevant in a very narrow region around 200-220 keV where the $^{14}$C
tail due to detector resolution smearing can extend up to 220 keV
(the end-point of the $\beta$-spectrum of $^{14}$C is 156 keV).
Above 220 keV the background is linear up to 400 keV as can be seen
from the experimental data (see Fig.4 from \cite{Borexino}). The
behaviour of the underlying background at E<200 keV doesn't influence
the result, because at these energies the total background, as noted
above, is dominated by the $^{14}$C beta decay. The assumption of
background linearity has been checked with MC simulations and is further
justified by the narrowness of the region relevant for the analysis
(about 30\% of the FWHM energy resolution). Ultimately, the choice
of the parent distribution is supported by the quality of the fit
$\chi^{2}=134.5/146$.\\
 It is worth pointing out that the number of parameters of the KKKT
fitting model (4 amplitudes for gaussian peaks, 2 parameters for the
linear background, and presumably parameters characterizing energy
calibration and linewidths) is at best the same as the number used
in the Borexino fit. Besides, the larger the number of parameters
the weaker the limit, so the number of parameters itself is not an
\textit{apriori} problem of the model\textit{.}\\

\item \textit{Strong and perhaps not unique cuts have been applied to reduce
the contamination of the spectrum in the range of interest by betas
and gammas from $^{40}$K and from $^{14}$C''.}\\
 It is not clear what is meant by the expression {}``perhaps not
unique cuts''. Cuts are never {}``unique'', but provided their
effect is well understood and characterized, they are a perfectly
valid technique for data analysis. The efficiency and effect of every
cut used in the CTF-II was thoroughly investigated.\\
 The most efficatious cut rejected events which reconstructed outside
of the inner, fiducial, part of the detector, exploiting the excellent
position sensitivity of the CTF detector. This technique of {}``active
shielding'' is widely used with high purity detectors. The fiducial
cut provides strong suppression of the external background, mainly
due 1.46\,MeV $\gamma$-rays emitted by the $^{40}$K contamination
in the strings of the nylon vessel hold-down system. $\beta$'s from
$^{40}$K do not contribute to the background because they do not
penetrate into the liquid scintillator.\\
 At the same time, we note that the $\beta$-background from $^{14}$C
decay, which is uniformly distributed throughout the detector's volume,
has the exact same spatial distribution as the signal of interest
which consists of $\gamma$-rays of 256\,keV.\\
 Thus, the relative reduction of the $^{14}$C signal due to the fiducial
volume cut can be used to define the fiducial volume of the detector.
This is another advantage of CTF setup in comparison with an HPGe
detector: the fiducial volume of the detector is determined experimentally
rather than by MonteCarlo methods.\\
 Another very effective cut removed events tagged by the muon identification
system, as is done in most low-background setups, including typical
HPGe detectors.\\
 The KKKT emphasis on the advantage of using {}``raw'' data is thus
at best unclear. 
\item \textit{{}``It is not clear that threshold effects on the spectrum
in the range of interest are really excluded''.}\\
 The threshold effect was really excluded, and it was explained in
detail in the Borexino article \cite{Borexino}. Anyway, threshold
effects would only influence the quality of the $^{14}$C fit near
the (software defined) threshold, and not the final result obtained
in the energy window which is well beyond the threshold. Once again,
the quality of the overall fit confirms the absense of a threshold
effect. 
\item \textit{{}``There do not exist direct measurements of the dependence
between light yield of the electrons and their energy for the scintillator
used in Borexino''.}\\
 The light yield was defined using the $^{14}$C end-point with high
precision using the model described in the text of the paper \cite{Borexino},
and was verified at higher energies using $^{40}$K 1.46 MeV $\gamma-$peak
and 514 keV $\gamma$'s from $^{85}$Kr decay into the excited state
of $^{85}$Rb, as explained in the Borexino article. There was no
need for additional measurement of light yield for the scintillator
used in Borexino as {}``self-calibration'' using dissolved natural
radioacive isotopes is the best imaginable calibration method.\\
 As a further check, the dependence of the lifetime limit on the scintillator
quenching factor was performed and presented in the Borexino paper.
The choice of any quenching factor in the realistic range from 0.01
up to 0.02 cm$^{-1}$MeV$^{-1}$ changes the resulting limit by no
more than 4\%. 
\item \textit{{}``The energy resolution in the Borexino experiment is by
a factor of 30 worse than that of the present Ge experiment''.}\\
 This statement is misleading since the sensitivity of different experimental
setups is not defined solely by their resolution. The ratio of sensitivities
of two detectors, I and II, for the signal of interest can be estimated
as:\\
 \[
\frac{\tau_{I}}{\tau_{II}}=\frac{\epsilon_{I}}{\epsilon_{II}}\sqrt{\frac{T_{I}}{T_{II}}\frac{N_{e_{I}}}{N_{e_{II}}}\frac{B_{II}}{B_{I}}\frac{\sigma_{II}}{\sigma_{I}}},\]
 where $\epsilon$ is detection efficiency, $T$ is data collection
time, $N_{e}$ is number of electrons, $B$ is specific background
and $\sigma$ is the detector's resolution. As one can see, the ratio
of the number of electrons (or the detector mass) and the ratio of
specific backgrounds have the same importance as the ratio of resolutions.
Borexino has much higher mass, and a much better specific background.\\
 In addition, the value quoted as the ratio of the resolutions ({}``a
factor of 30'') is also misleading. As explained in the KKKT article,
the resolutions of interest are not the intrinsic energy resolutions
of the detectors, but the Doppler broadened resolutions: at least
8 keV instead of 2.4 for the HPGe detector. The Doppler-broadened
resolution of the Counting Test Facility is only 9 times worse than
the HPGe counterpart. Moreover, a realistic comparison of sensitivity
should contain the width weighted over all the electron shells. 
\end{enumerate}

\section*{Final remarks}

The final result stated by the KKKT authors is 5 times better than
their own estimate of the detector's raw sensitivity using the \char`\"{}1
$\sigma$ method\char`\"{}, while Borexino's result is about 2 times
worse than its raw sensitivity that can be obtained using the same
\char`\"{}1 $\sigma$ method\char`\"{} (this sensitivity was not given
in the Borexino article, but the value can be extracted from the data
presented). So, relative to the KKKT result, the Borexino result stands
on even more conservative premises and accurately reflects its systematic
uncertainties.

On a technical note, the result obtained by the Borexino collaboration
is quoted at the 90\% c.l.. In contrast, the result from KKKT is quoted
at the 68\%~c.l.. However, in the {}``Conclusions'' of KKKT the
two numeric values are directly compared, without explicit mention
of the difference in the confidence levels. Finally, the best KKKT
result considers only outer shell electrons, calling into question
their assumed number of candidate electrons.

We conclude that the criticism of the Borexino result in KKKT is unsubstantiated.
Moreover, the {}``indication of a signal on 1.4 $\sigma$'' purported
in KKKT is excluded by results from both the Borexino~\cite{Borexino}
and DAMA~\cite{DAMA} collaborations.

\end{document}